\documentclass[prb,twocolumn,showpacs,superscriptaddress,floatfix]{revtex4}
\usepackage{graphicx}
\usepackage{float}

\begin{document}

\title{Wavelength de-multiplexing properties of a single aperture
flanked by periodic arrays of indentations}
\author{J. Bravo-Abad}
\affiliation{\mbox{Departamento de F\'{\i}sica Te\'{o}rica de la Materia Condensada
, Universidad Aut\'onoma de Madrid, E-28049 Madrid, Spain}}
\author{F.J. Garc\'{\i}a-Vidal}
\affiliation{\mbox{Departamento de F\'{\i}sica Te\'{o}rica de la Materia Condensada
, Universidad Aut\'onoma de Madrid, E-28049 Madrid, Spain}}
\author{L. Mart\'{\i}n-Moreno}
\affiliation{\mbox{Departamento de F\'{\i}sica de la Materia Condensada, ICMA-CSIC,
 Universidad de Zaragoza, E-50009 Zaragoza, Spain}}

\begin{abstract}
% Text of abstract
In this paper we explore the transmission properties of single
subwavelength apertures perforated in thin metallic films flanked
by asymmetric configurations of periodic arrays of indentations.
It is shown how the corrugation in the input side can be used to
transmit selectively only two different wavelengths. Also, by
tuning the geometrical parameters defining the corrugation of the
output side, these two chosen wavelengths can emerge from the
structure as two very narrow beams propagating at well-defined
directions. This new ability of structured metals can be used as a
base to build micron-sized wavelength de-multiplexers.
\end{abstract}



\maketitle
% main text
\section{Introduction}
\label{} The new technological abilities for structuring metals in
the micro and nanometer length scales are triggering a renewed
interest in the optical properties of metals
\cite{Sigalas95,Barnes96,Garcia96,Fan96,Sievenpiper98,Pendry99,Fleming02}.
A breakthrough in the field was the experimental discovery
\cite{Ebbesen} that the optical transmission through an array of
sub-wavelength holes is, for certain wavelengths, boosted orders
of magnitude over what a theory of independent holes would
predict.

In subsequent works it was shown how the transmission through a
{\bf single} aperture can be boosted if it is appropriately
flanked by surface corrugations in the side light is impinging on
\cite{Grupp99,Thio,Lezec}. Corrugations on the exit surface
practically do not modify the {\it total} transmittance but, for
some resonant wavelengths, greatly modify the angular distribution
of the transmitted light, leading to strong beaming effects
\cite{Lezec,LMM03}. This was shown for a hole surrounded by
corrugations in the so-called bull's eye geometry and for a single
slit flanked by one-dimensional grooves. It has been shown that
these two processes (enhanced transmission and beaming) work
independently \cite{LMM03,FJ03}, so it is possible to have more
light passing through an aperture and have it transmitted in a
narrow beam by patterning both the input and exit sides. These
effects are associated to the formation of electromagnetic
resonances (leaky modes) running along the metal surface, and
radiating as they move due to the coupling to homogeneous waves in
vacuum. The spectral position of these EM resonances depend on the
geometrical parameters defining the corrugations (distance between
indentations, and their widths and depths).

These studies were done in a symmetric configuration of a finite
number of grooves surrounding the central slit. In this paper we
show that, to a large extend, the properties of the groove arrays
at the left and right of the slit behave independently. This
property adds to the interpretation of the properties of the
system in terms of running surface waves. It also allows, through
the appropriate choice of geometrical parameters for left and
right groove arrays at both entrance and exit surfaces, this type
of structures to act as wavelength de-multiplexers: the input side
of the device can be used as a filter for two particular
wavelengths and the corrugations on the exit side can be tailored
in order to beam each of these two chosen wavelengths in different
directions.

The paper is organized as follows: in Section 2 we present the
theoretical formalism. Section 3 presents the transmission results
for a single slit with an array of groove to one side, while the
transmission angular distribution is considered in Section 4. In
Section 5, we show how this kind of structures may be used for
wavelength demultiplexing (WDM).

\section{Theoretical formalism}

We are interested in analyzing the transmission properties of the
system depicted in Fig.1: a single slit flanked by finite arrays
of one-dimensional rectangular indentations. The central slit may
be flanked by arrays of grooves at both input and output metal
surfaces and, in each case, the arrays may be different to the
left and right of the slit. We refer to these four arrays as IL,
IR, OL and OR (see Fig.1), depending on whether they are
input-left, input-right, output-left and output-right,
respectively. Each groove ($\alpha$) is characterized by its
position ($r_\alpha$), its width ($a_\alpha$) and its depth
($h_\alpha$), while the slit (labelled by the sub-index 0) is
considered to be placed at the origin, having a width $a_0$. Metal
thickness is $W$.

\begin{figure}
\begin{center}
\includegraphics[width=\columnwidth]{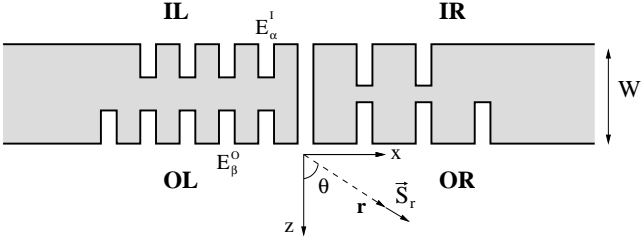}
\end{center}
\caption{Schematic illustration of the structure under study. IL,
IR, OL and OR label the input-left, input-right, output-left and
output-right arrays of grooves, respectively. The reference system
considered is also plotted.}
\end{figure}

Let us concentrate in the case of p-polarized light for which
interesting resonant effects have been reported. Electromagnetic
(EM) fields associated to this scattering process are expanded by
a set of plane waves in vacuum and by a modal expansion inside the
indentations (slit and grooves). For this last case, as we are
mainly interested in sub-wavelength apertures, we consider that
only the fundamental propagating eigenmode is excited. Then, for
$x$ inside indentation $\alpha$, the wavefield is a linear
combination of $\phi_\alpha(x)\exp(\pm \imath kz)$, where
$k=2\pi/\lambda$ and $\phi_\alpha(x)=1/\sqrt{a_\alpha}$. We assume
that the metal behaves as a perfect metal, expelling the electric
field from its bulk. One advantage of the perfect metal
approximation is that as all length scales associated to the
dielectric constant (skin depth and absorption length) disappear
from the problem, all results are directly exportable to other
frequency regimes (microwave or infrared regimes), simply by
scaling appropriately the geometrical parameters defining the
structure. With respect to the applicability of this approach in
the optical regime, we have demonstrated in previous theoretical
works \cite{FJ02} that this is a reasonable approximation when
analyzing optical properties of nanostructured metals like silver
or gold in simpler systems like reflection gratings or 1D arrays
of subwavelength slits. Moreover, within this framework, we have
been able to reproduce \cite{LMM03,FJ03} in semi-quantitative
terms the reported phenomena of enhanced transmission and beaming
appearing in the optical regime for finite structures.

In order to extract the expansion coefficients in vacuum (related
to the reflection and transmission functions) and indentations
regions, we match the parallel components of the EM fields ($E_x$
and $H_y$) at the interfaces present in the system. By projecting
the matching equation for $E_x$ at $z=-W$ and $z=0$ onto plane
waves, we can express the reflection and transmission coefficients
in terms of the wavefield expansion coefficients inside the
indentations. $H_y$ must be continuous only at the indentation
openings and also (as we are assuming perfect metal boundary
conditions) $E_x=0$ at the bottom of the grooves. Therefore, all
matching equations can be expressed as a function of the set of
expansion coefficients inside the indentations:
[${E^{I}_{\alpha},E^{O}_{\beta}}$](see Fig.1). The sub-set
($E^{I}_{\alpha}$) give the $x$-component of the electric field at
$z=-W^+$ through the equation: $E_x(x,z=-W^+)=\sum_\alpha
E^{I}_{\alpha} \phi_\alpha(x)$, where $\alpha$ runs from
$-N^{I}_{L}$ to $+N^{I}_{R}$ ($N^{I}_{L}$ and $N^{I}_{R}$ are the
number of grooves to the left and right of the slit in the input
side, respectively). ($E^{O}_{\beta}$) is related to the
$x$-component of the electric field at the exit interface
$z=0^-$:$E_x(x,z=0^-)=\sum_\beta E^{O}_{\beta} \phi_\beta(x)$,
$\beta$ running from $-N^{O}_{L}$ to $+N^{O}_{R}$ ($N^{O}_{L}$ and
$N^{O}_{R}$ are the number of grooves to the left and right of the
slit in the output side, respectively). In both expansions,
sub-index $0$ stems for the central slit. After some elementary
algebra, it is possible to find the set of
$(N^{I}_{L}+N^{I}_{R}+N^{O}_{L}+N^{O}_{R}+2)$ equations that
govern the behavior of [${E^{I}_{\alpha},E^{O}_{\beta}}$]:

\begin{eqnarray}
\noindent \left[ G_{\alpha \alpha} - \epsilon_{\alpha}  \right]
E^{I}_\alpha + \sum_{\gamma \neq \alpha} G_{\alpha \gamma}
E^{I}_\gamma -G_v E^{O}_{0}\delta_{\alpha 0}&=&I_\alpha \\
\nonumber \left[ G_{\beta \beta} - \epsilon_{\beta}  \right]
E^{O}_\beta + \sum_{\eta \neq \beta} G_{\beta \eta} E^{O}_\eta
-G_v E^{I}_{0}\delta_{\beta 0}&=&0
\end{eqnarray}

The first set of equations accounts for the expansion coefficients
at the input interface and the sub-indexes $\alpha$ and $\gamma$
run from $-N^{I}_{L}$ to $N^{I}_{R}$. Due to the normalization we
use for the transmittance (see below), the illumination term
$I_\alpha$ is equal to 2 for all $\alpha$.  The second set of
equations is homogeneous (the output interface is not directly
illuminated), and $\beta$ and $\eta$ ranges from $-N^{O}_{L}$ to
$N^{O}_{R}$. Note that the only connection between the two
interfaces is via the central slit and appears through the term
$G_v E^{O}_{0}\delta_{\alpha 0}$ in the first set and by $G_v
E^{I}_{0}\delta_{\beta 0}$ in the second one. In Refs.
\cite{LMM03,FJ03} we explain in detail the origin and values of
the different terms appearing in Eq.(1). For instance, $G_{\alpha
\beta}E_\beta$ is the contribution to the $E_x$ field at
indentation $\alpha$ coming from light scattered at indentation
$\beta$. In fact, $G_{\alpha \beta}$ is the projection onto
wavefields $\phi_{\alpha}$ and $\phi_{\beta}$ ($G_{\alpha \beta}
=< \phi_{\alpha}\mid G \mid \phi_{\beta}>$) of the Green's
function: $G({\bf r},{\bf r^\prime})=\frac{\imath \pi}{\lambda}
H_{0}^{(1)}(k\mid{\bf r}-{\bf r^\prime}\mid)$, $H_0^{(1)}$ being
the Hankel function of the first kind. $G_{\alpha \alpha} =<
\phi_{\alpha}\mid G \mid \phi_{\alpha}>$ is a self-interaction
that takes into account the coupling of the eigenmode at
indentation $\alpha$ with the radiative modes in vacuum regions.
This quantity depends only on the ratio between indentation width
and wavelength of light, $a_{\alpha}/\lambda$. As said before, in
this type of structures both sides of the corrugated metal film
are only coupled by the term $G_v$ that is a function of metal
thickness: $G_v=1/\sin (kW)$. The diagonal terms $\epsilon_\alpha$
are related to the back and forth bouncing of the EM-fields inside
indentation $\alpha$: $\epsilon_\alpha=\cot(kh_{\alpha})$ for
$\alpha\neq0$ and $\epsilon_0=\cot (kW)$ for the central slit.
Once the values for [{$E^{I}_{\alpha},E^{O}_{\beta}$}] are
calculated, the total transmittance is a function of the
amplitudes of the E-field at the entrance and exit of the central
slit: $T= Im \left\{ E^{I}_0 E^{O*}_0 \right\}/ \sin (kW)$.

Also the EM fields in real space can be expressed in terms of the
set [$E^{I}_{\alpha},E^{O}_{\beta}$]. For example, in the
transmission region ($z > 0$):

\begin{equation}
H_y(\vec{r}) = \frac{1}{\mu_0 c} \sum_\alpha E^{O}_\alpha
G(\alpha, \vec{r})
\end{equation}
all other components of the EM field can be obtained from
$H_y(\vec{r})$ for the polarization considered. Here $ G(\alpha,
\vec{r}) = \imath k / 2 \int \phi_\alpha^*(x) H_0^{(1)}(k |x
\vec{u}_x - \vec{r}|) \, dx $. Similar expressions can be obtained
for the EM-fields in the reflection region and within the
indentations.

\section{Transmission properties}

In Ref. \cite{FJ03} we analyzed, in a symmetric groove
configuration, the conditions for having large transmittance, i.e.
for having large E-field at the central slit. We identified three
main resonant mechanisms: slit waveguide mode excitation
(controlled by the metal thickness), groove cavity mode
(controlled by the groove depths) and in-phase groove re-emission
processes (that is maximum for wavelengths of the order of the
period of the array, $d$). Large enhancements in the transmission
for resonant wavelengths can be obtained when the three mechanisms
coincide; this means that if we fix $d$ there is an optimum value
of the depth of the grooves in which the optical transmission is
maximum.

In this section we present results for the total transmittance $T$
of a system composed of a slit and, in the most general case, four
periodic finite arrays of grooves. We normalize the transmittance
to the incident flux in the direction normal to the metal surfaces
and to the slit area. In this way $T$ for point particles would be
$1$. Even with the simplifications introduced, the structure under
study is still characterized by a large number of geometrical
parameters. In this paper all indentation widths ($a$) are equal
and the grooves arrays, when present, have all the same number of
grooves ($N$). All results presented here are for $a=40 nm$,
$N=10$ and metal thickness $W=200$nm, which are typical values
considered previously in related experiments (later on we will
explain the reason for choosing this particular metal thickness).

For WDM it is convenient to have at least two wavelengths
transmitted primarily through the structure. Let us select,
arbitrarily and simply for proof of principle purposes,
$\lambda_1=500$nm and $\lambda_2=600$nm. We consider first the
even simpler system of a single slit flanked by just one groove
array in the input side (IL). For the chosen values of $a$ and
$N$, after an optimization process of $d_L$ and $h_L$ we find
enhanced transmission at $\lambda_1$ for $d_L= 470$nm,$h_L=75$nm.
For this set of parameters, groove cavity mode excitation and
in-phase groove re-emission mechanisms are tuned to appear at
$\lambda_1$. The curve $T(\lambda)$ for this case is rendered in
Fig. 2 (blue curve), showing clearly the transmission resonance at
$\lambda=500$nm due to the presence of grooves (compare this curve
with the black curve of Fig. 2, showing $T(\lambda)$ for a bare
single slit). The green curve of Fig.2 presents the $T(\lambda)$
for a similar system, also with only one groove array next to the
single slit but this time with lattice parameter $d_R=560$nm and
groove depth $h_R=95$nm in order to have resonant transmission at
$\lambda=600$nm. Also in Fig. 2 (red curve) we show $T(\lambda)$
for the case of a single slit flanked by the two groove arrays
considered previously. This last curve shows that the transmission
resonances associated to each of the arrays survive, although
somehow modified, the close presence of the other array. The fact
that when both arrays are present the transmission peaks at
$\lambda=500$ and $\lambda=600$nm have the same height is neither
casual nor reflects any hidden property of the system: simply the
value of the metal thickness ($W=200$nm) has been tuned for this
to be the case. Notice that similar transmission peak heights for
the selected wavelengths is a bonus for WDM, and we just wanted to
show that this condition may be fulfilled in this kind of
structures. The case presented here is, however, not necessarily
optimal. Probably it is possible to make the selected wavelengths
to profit from slit waveguide resonances
\cite{Porto99,Collin,Takakura,Sambles02} (the third enhancing
transmission mechanism previously discussed) and get even larger
transmittances. The inset to Fig. 2 shows that in asymmetric
structures, as previously found for symmetric ones \cite{FJ03},
$T(\lambda)$ is mostly independent of the patterning of the output
surface.

\begin{figure}
\begin{center}
\includegraphics[width=\columnwidth]{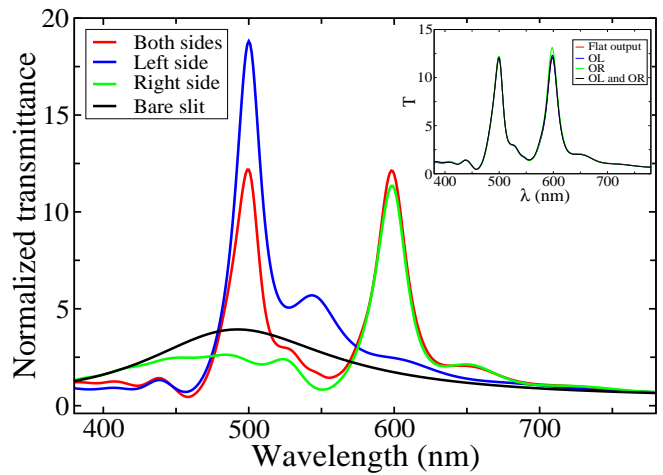}
\end{center}
\caption{Normalized-to-area transmittance plotted as a function of
the wavelength for a structure defined by a flat exit surface and
a input surface with grooves at both sides (red line), left side
(blue line) and right side (green line). Black line corresponds to
a single slit without corrugations. Inset shows the results
considering grooves at both sides of the input surface and
different configurations of the output surface.}
\end{figure}

\section{Far-field angular distribution of transmitted light}

In this section we present results for the far-field angular
dependence of the radial component of the Poynting vector,
$S_r(\vec{r})$, evaluated in the output region. In the
subwavelength limit, the transmission angular distribution of this
kind of systems depends only on the output corrugation, up to a
constant that accounts for the transmittance at that wavelength
\cite{LMM03}. Therefore, it is convenient to define the far-field
angular transmission distribution $I(\theta) \equiv r
S^{nor}_r(\theta)$ ($ r \gg N d, \lambda$), where
$\vec{S}^{nor}(\lambda, \vec{r})= \vec{S}(\lambda,
\vec{r})/T(\lambda)$. Recent studies in the optical
\cite{Lezec,LMM03} and microwave \cite{Hibbins02} regimes have
found that $I(\theta)$ for a single slit flanked by grooves arrays
in the output side shows beaming effects for certain frequency
ranges. In Ref.\cite{Lezec} a phenomenological model, based on the
hypothesis of the existence of a leaky wave running along the
surface, was enough to explain the beaming phenomena. A subsequent
first-principle model \cite{LMM03} supported the basics of the
phenomenological approach and the origin of the leaky surface
wave, and was in good agreement with the experimental data.
Basically, beaming can be understood as follows: the EM field
leaving the slit couples to a running surface wave (which is the
result of self-consistent wandering back and forth of EM fields at
the grooves, which act as surface scattering centers); this
surface wave accumulates phase as it travels along the surface and
it radiates to vacuum at the groove positions. So, for emission
purposes, the groove array behaves as an effective diffraction
grating, but illuminated at a (wavelength dependent) angle, which
is related to the phase accumulated by the surface running wave.
Actually, the emission intensity is different at the different
radiating points of this equivalent diffraction grating, and must
be calculated self-consistently from Eq.(1). However the final
result shows that, at some angles, the extra optical path of waves
leaving from different grooves cancel the phase of the running
wave, resulting in a maximum in $I(\theta)$.

Both studies \cite{Lezec,LMM03} previously referred to considered
a symmetric distribution of grooves around the central slit and,
therefore, showed a symmetric $I(\theta)$. In this paper we
consider asymmetric structures, showing that beaming occurs even
when a slit is flanked by just a single groove array. Also we show
that, to a large extend, the beam produced by a groove array is
not modified if a different groove array is placed also at the
output metal surface but at the other side of the slit. This is
illustrated in Fig.3 where panel (a) shows $I(\lambda,\theta)$ for
a system with a single groove array is placed in the OL
configuration with geometrical parameters $d_L=470$nm and
$h_L=75$nm. Notice that, in section 3, we found that this set of
parameters are optimum in order to have enhanced transmission at
$\lambda_1=500$nm. Fig. 3a shows that this asymmetric structure
presents beaming for a range of wavelengths and that maximum
beaming appears at the same resonant wavelength $\lambda_1$. It is
worth noticing that even in this asymmetric case, maximum beaming
occurs at $\theta=0^0$. A similar representation appears in panel
(b), this time for an array of grooves in OR configuration with
$d_R=560$nm and $h_R=95$nm which optimizes both transmission and
$I(\theta=0^0)$ for $\lambda_2=600$nm. Panel (c) renders
$I(\lambda,\theta)$ when both OL and OR arrays previously
considered are present in the structure. The overall picture, more
precisely lines in the ($\lambda-\theta$) plane showing large
values of $I(\lambda,\theta)$ corresponds to the addition of
beaming lines of panels (a) and (b). However, panel (c) also shows
that radiation coming from left and right groove arrays is not
totally independent. Instead, there is a small shift in both
wavelengths at which maximum beaming appears and the associated
angles which now may move away from normal direction. This last
fact will be of paramount importance in order to use this kind of
structures as wavelength de-multiplexers.

\begin{figure}
\begin{center}
\includegraphics[width=\columnwidth]{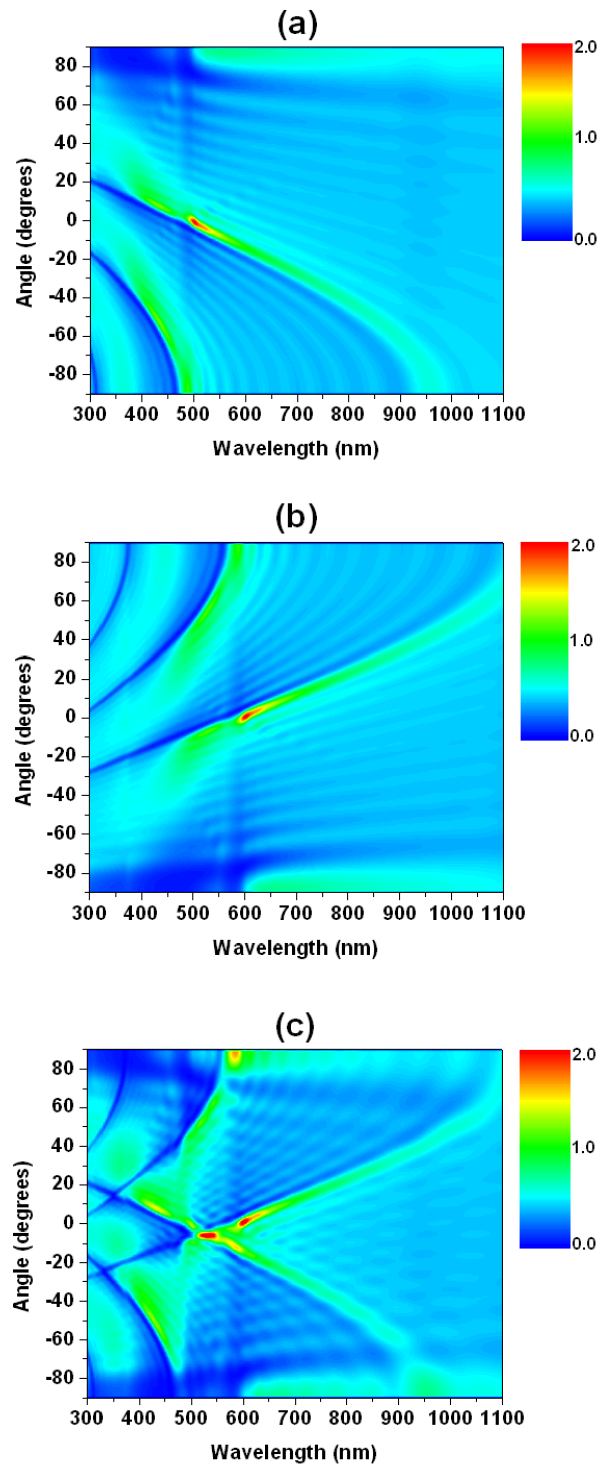}
\end{center}
\caption{Contour plots of $I(\lambda,\theta)$ with $\theta$
defined in Fig. 1. Three different configurations of the output
surface are considered. (a) OL with $h$=75 nm and $d$=470 nm. (b)
OR with $h$=95 nm and $d$=600 nm. (c) OL+OR with the same
geometrical parameters as in (a) and (b).}
\end{figure}

\section{Wavelength de-multiplexer device}

The two abilities of these systems described in the two previous
sections, namely i) the selective transmission of two wavelengths
when structuring the input side of the metallic film (Section 3)
and ii) beaming at the same two wavelengths in different
well-defined directions (Section 4) can be combined to propose a
new application of single apertures surrounded by finite arrays of
grooves.: this kind of structures can be used as wavelength
de-multiplexers of micro-meter dimensions. This new functionality
is illustrated in Fig. 4 where $rS_r(\theta)$ in the far-field
region is calculated for different wavelengths for a structure in
which four groove arrays are present. Geometrical parameters of
these arrays are essentially those considered in Fig.2(red curve)
and Fig.3c, only that one of the grooves array in the input side
has been slightly modified in order to obtain enhanced
transmission at $\lambda=530$nm. For the two resonant wavelengths
($\lambda=530$nm and $\lambda=600$nm) light emerges in the form of
very narrow beams presenting very low divergence (around $5^0$)
and can be collected at an angle of $6^0$ for $\lambda=530$nm and
$0^0$ for $\lambda=600$nm. Note that due to the interplay between
the left and right grooves arrays in the output side, light of
wavelength $600$nm can also be found at a larger collection angle
of $14^0$. For the rest of wavelengths (see Fig.4) optical
transmission is extremely low and its angular distribution is
nearly uniform.

\begin{figure}
\begin{center}
\includegraphics[width=\columnwidth]{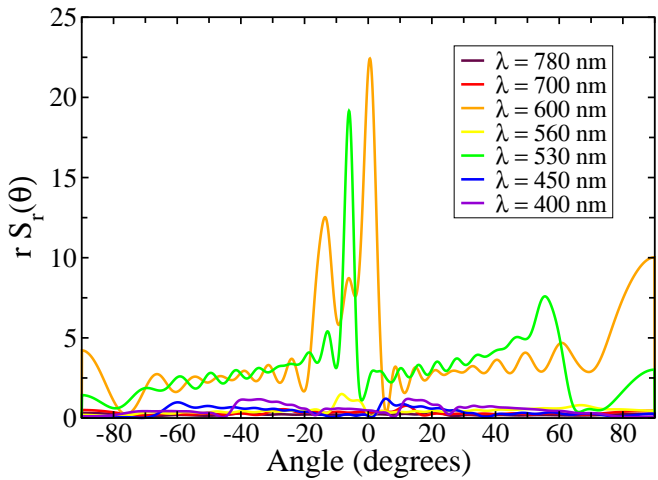}
\end{center}
\caption{Dependence of $r·S_r(\theta)$ (see text) on the angle for
several wavelengths. The result of each wavelength is plotted with
its corresponding color in the visible spectrum. Notice beaming
effect present for orange and green light.}
\end{figure}

To further illustrate the resonant nature of this phenomenom, in
Fig. 5 we show the near-field pictures (E-field intensity) in the
output region corresponding to the transmission processes for the
two resonant wavelengths ($\lambda=530$nm and $\lambda=600$nm) and
for a non-resonant case ($\lambda=700$nm). For this wavelength,
grooves are not playing any role; E-field intensity is low and the
transmission angular distribution is rather uniform. However in
the first two cases, as clearly seen in panels (a) and (b), light
is emitted not only by the central slit but also by the grooves
located at left and right sides of the aperture in the output
side. As a result of the interference between these re-emission
processes narrow beams of light emerge from the structure and are
already present in the near-field regime. Fingerprints of the
running surface waves that are at the origin of the beaming
phenomena can also be visualized in these pictures.

\begin{figure}
\begin{center}
\includegraphics[width=\columnwidth]{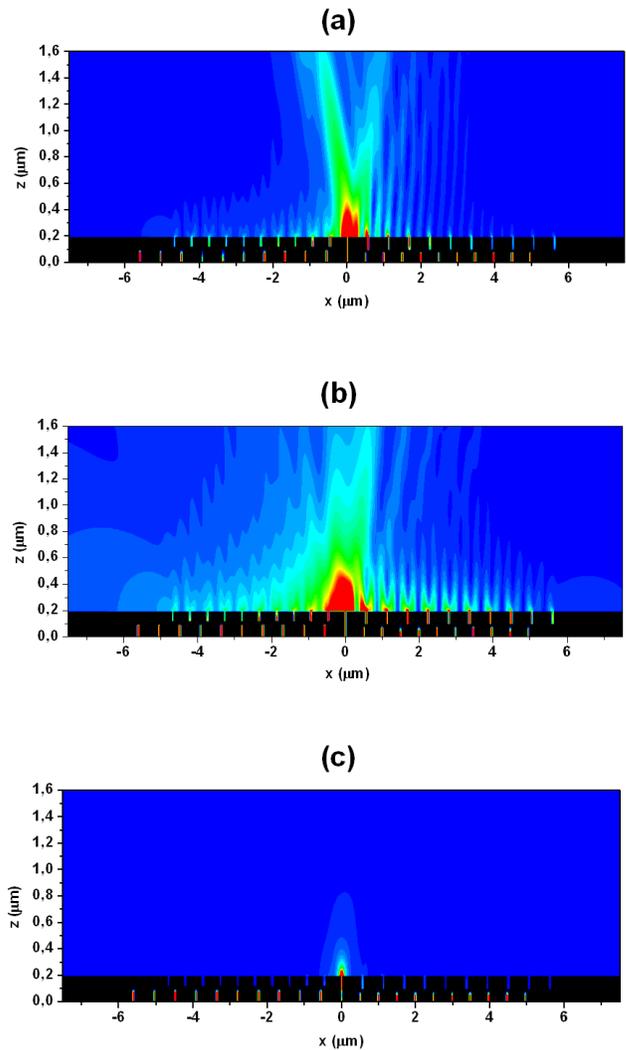}
\end{center}
\caption{Contour plots of the electric field intensity for the
proposed demultiplexing structure. Three different wavelengths are
considered. (a) $\lambda$=530 nm. (b)$\lambda$=600 nm.
(c)$\lambda$=700 nm (off-resonance). Red (blue) corresponds to
maximum (minimum) intensity. Black areas show the metallic
regions.}
\end{figure}

In conclusion, we have shown how a single subwavelength aperture
surrounded by finite arrays of indentations may be used as a
wavelength de-multiplexer. The two selected wavelengths can be
chosen by just modifying the geometrical parameters defining the
structure (period of the arrays at left and right of the central
slit and widths and depths of the indentations). A practical
advantage of this type of {\it plasmon} wavelength de-multiplexer
device with respect to others based on, for example, photonic
crystals \cite{Noda} is that light emerging from our proposed
device forms two collimated beams presenting both a very low
divergence. This divergence can also be controlled and reduced by
increasing the number of indentations in the output side.
Implementation of this device to build up a multi-frequency
de-multiplexer can be achieved by just adding up other single
apertures and the corresponding indentations with the appropriate
geometrical parameters to the structure analyzed in this
communication. The fact that input and output corrugations seem to
act independently and also that the interplay between left and
right indentations is weak allows us to state that this type of
structured metals could be used as building-blocks of several
opto-electronic devices of reduced dimensions.

\section*{Acknowledgements}

Financial support by the Spanish MCyT under grant BES-2003-0374
and contracts MAT2002-01534 and MAT2002-00139 is gratefully
acknowledged.

% \appendix

% \section{}
% \label{}

\end{document}